\documentclass[12pt]{JHEP3}

\usepackage{amsmath}
\usepackage{amssymb}
\usepackage{cite}
\usepackage{array}

\usepackage[german,american]{babel}
\usepackage[latin1]{inputenc}

\usepackage{fancyhdr} 

\fancypagestyle{plain}{%
  \fancyhf{}                       
  \cfoot{\thepage}                 
  
  }

\newcommand{\ol}[1]{\overline{#1}}

\newcommand{\beqs}{\begin{equation*}}
\newcommand{\beq}{\begin{equation}}

\newcommand{\eeqs}{\end{equation*}}
\newcommand{\eeq}{\end{equation}}

\newcommand{\beqas}{\begin{eqnarray*}}
\newcommand{\beqa}{\begin{eqnarray}}

\newcommand{\eeqas}{\end{eqnarray*}}
\newcommand{\eeqa}{\end{eqnarray}}

\newcommand{\eps}{\varepsilon}
\newcommand{\al}{\alpha}
\newcommand{\be}{\beta}
\newcommand{\ga}{\gamma}
\newcommand{\de}{\delta}
\newcommand{\om}{\omega}

\newcommand{\la}{\lambda}

\newcommand{\Si}{\Sigma}

\newcommand{\blist}{\begin{itemize}}

\newcommand{\elist}{\end{itemize}}

\providecommand{\href}[2]{#2}

\newcommand{\pd}[2]{\frac{\partial #1}{\partial #2}}
\newcommand{\pld}[2]{\frac{\partial^L #1}{\partial #2}}

\newcommand{\td}[2]{\frac{\mathrm{d} #1}{\mathrm{d} #2}}
\newcommand{\lrpd}[1]{\overleftrightarrow{\partial_{#1}}}

\newcommand{\poiss}[2]{\{ #1 , #2 \}}
\newcommand{\spoiss}[2]{ \{ #1 , #2 \}}
\newcommand{\dirac}[2]{\{ #1 , #2 \}^*}

\newcommand{\chib}{\overline{\chi}}
\newcommand{\chid}{\chi^\dagger}

\newcommand{\plr}[1]{\overleftrightarrow{\partial_{#1}}}

\author{René Meyer\\ Institute for Theoretical Physics, University of Leipzig\\ Augustusplatz 10-11, D-04109 Leipzig, Germany\\ E-mail: \email{rene.meyer@itp.uni-leipzig.de}}

\abstract{Generalized Dilaton Theories in two dimensions coupled to
  Dirac fermions are subjected to constraint analysis. Three first
  class secondary constraints are found, corresponding to one local
  Lorentz symmetry and two diffeomorphisms. Moreover, the system also
  yields second class constraints from the fermions. The algebra of
  first class constraints is calculated in some detail, and is found
  to be related to the classical Virasoro algebra.}

\title{Constraints in two-dimensional Dilaton Gravity with Fermions}

\begin{document}

\section{Introduction}\label{sec:intro}

Generalized Dilaton Theories (GDTs) in two dimensions arise from
several fields of physics (for an exhaustive review, see
\cite{Grumiller:2002nm}). Prominent examples would be spherical
reduced Einstein-Hilbert gravity from $D>2$ dimensions, the ``string
inspired'' dilaton model (CGHS model, \cite{Callan:1992rs}) or the
models put forward by Jackiw \& Teitelboim
\cite{Jackiw:1984,Teitelboim:1984} and Katanaev \& Volovich
\cite{Katanaev:1986wk} in the 1980s. All these models can be subsumed
into one action \cite{Russo:1992yg,Odintsov:1991qu,Banks:1991mk}
\beq\label{dilatonaction}
S^{(GDT)} = \int d²x \sqrt{-g} \left[\frac{R}{2}X - \frac{U(X)}{2}(\nabla X)²+V(X) \right]
\eeq
with $X$ being the dilaton field, $R$ the Ricci scalar and $U(X)$ and
$V(X)$ arbitrary potentials specifying the model.

It turns then out that there exists a classically equivalent
formulation of \eqref{dilatonaction} in terms of Cartan variables,
namely the Vielbein $e^a=e^a_\mu dx^\mu$, the spin connection
$\omega^{ab}=\omega_{\mu}^{ab}dx^\mu=:\omega\epsilon^{ab}$; the dilaton $X$ and
additional auxiliary fields $X^\pm$. This First Order Gravity (FOG)
action reads \cite{Schaller:1994es}
\beq\label{FOGaction}
S_{FOG} = \int\limits_{{\cal M}_2} \left[ X^a(De)_a+Xd\omega+\epsilon{\cal V}(X^+X^-;X)\right]
\eeq
Here we allow also for non-vanishing torsion $T^a = (De)^a$ terms, and
the proof of classical equivalence (for potentials ${\cal V}(X^+X^-;X)
= U(X) X^a X_a + V(X)$), which can be found in ch. 2 of
\cite{Grumiller:2002nm}, amounts to using the equations of motion for
the $X^a$ to eliminate the torsion dependent part of the spin
connection.

The FOG action is our starting point for the analysis presented.  We
will explain the coupling to Dirac fermions in sec.
\ref{sec:fermions}. In Sec. \ref{sec:analysis} we will analyse the
constraints of the theory and obtain the constraint algebra. Sec.
\ref{sec:witt} relates this algebra with the well known classical
Virasoro algebra (or Witt algebra) in two dimensions, and Sec. 5
contains some discussion of the main results. The conventions used in
this article can be found in Appendix \ref{app:A}. Appendices
\ref{app:B} and \ref{app:C} contain equations necessary for proving
the main results of this paper.

It should be noted that a similar analysis has been carried out for
scalar fields coupled to FOG in \cite{Grumiller:2001ea} and for
massless, not self interacting and minimally coupled fermions
\cite{wal01}. We extend this analysis to the general case of massive,
self-interacting Dirac fermions with non-minimal coupling. (What is
meant by (non-)minimal coupling will be explained in the next
section.) One of the first works considering GDTs (with $U(X)=0$) coupled to fermions  was \cite{Cavaglia:1998yp}.
Even before, specific models were used as toy
models for Black Hole evaporation \cite{Nojiri:1992st,Ori:2001xc} and
more recently in a paper by Thorlacius et. al.  \cite{Frolov:2005ps}
For another remark on this, see also Sec.  \ref{sec:conclusions}.

\section{Fermions}\label{sec:fermions}

We add to \eqref{FOGaction} an action for the fermion fields
consisting of the well known kinetic term and a general self
interaction\footnote{Note that $\ast e^a \neq e^a$.}
\beqa\label{wholeaction}
S & = & S_{FOG} + S_{kin} + S_{SI}\\\nonumber
S_{kin} & = & - \frac{i}{2} \int\limits_{{\cal M}_2} f(X)\; (*e^a) \wedge (\chib \ga_a \overleftrightarrow{\mathrm{d}}\chi)\\\label{fermionkinetic}
        & = & \frac{i}{2}\int d²x (e) f(X) {e_a}^\mu (\chib \ga^a \plr{\mu} \chi)\\ 
\nonumber
S_{SI}  & = & - \int\limits_{{\cal M}_2} \epsilon h(X) g(\chib\chi)\\\label{fermionSI}
        & = & \int d^2x (e) h(X) g(\chib\chi)
\eeqa

Here all possible boundary terms coming from the fermions have been
omitted. The functions $f(X)$ and $h(X)$ introduce a coupling to the
dilaton field. If both are constant, i.e. $f \propto h =
\mathrm{const.}$, we call the fermions minimally coupled, and
non minimally coupled otherwise.  Because spinors are anti commuting
Grassmann fields, a Taylor expansion of $g$ yields at at most a
quartic term, $g(\chib\chi) = g_0 + m\chib\chi + \la (\chib\chi)²$;
and the constant term can always be absorbed into $V(X)$.

Note that in two dimensions the kinetic term is independent of the
spin connection: In arbitrary dimension, the action for the kinetic
term would be\footnote{h.c. means hermitian conjugate}
\cite{Birrell:1989}
\beqs
\frac{i}{2} \int d^n x \det(e^a_\mu) \left[ f(X) e_a^\mu (\chib \ga^a (\partial_\mu + {\om_\mu}^{bc}\Si_{bc})\chi) + \mathrm{h.c.}\right]\eeqs
For $n=2$ however, there is only one independent generator of
Lorentz transformations $\Si_{01} = \frac{1}{4}[\ga_0,\ga_1]=-\frac{\ga_*}{2}$,
and with $\{\ga_a,\ga_*\}=0$ the terms in \eqref{fermionSI} containing the spin connection
vanish
\beqas
&   & -\frac{i}{4}f(X) (\ast e^a) \wedge \om \chid(\ga^0\ga_a\ga_* - \ga_*\ga_a^\dagger \ga^0)\chi\\
& = & -\frac{i}{4}f(X) (\ast e^a) \wedge \om \chid\underbrace{(\ga^0\ga_a - \ga_a^\dagger \ga^0)}\limits_{= 0} \ga_*\chi = 0
\eeqas

\section{Hamiltonian Analysis}\label{sec:analysis}

For the sake of better memorability, we henceforth denote the
canonical coordinates and momenta by
\beqas\nonumber
\ol{q}^i & = & (\om_0,e_0^-,e_0^+)\\
q^i      & = & (\om_1,e_1^-,e_1^+),\hspace{9.5mm} i=1,2,3 \\
p_i      & = & (X,X^+,X^-)\\\label{Qj}
Q^\al      & = & (\chi_0,\chi_1,\chid_0,\chid_1),\hspace{3mm} \al=0,1,2,3
\eeqas
The canonical structure on the phase space is given by Poisson
brackets
\beqa\label{poisson}
\spoiss{q^i(x)}{p_j(y)} & = & \de^i_j \de(x-y) \\ \nonumber
\spoiss{Q^\al(x)}{P_\be(y)} & = & - \de^\al_\be \de(x-y)
\eeqa
where the $P_\be$ are canonical momenta for the spinors, and not
present explicitly in \eqref{fermionkinetic} and \eqref{fermionSI}.

\subsection{Primary and Secondary Constraints}

In our system there occur both primary first and second class
constraints. A look at the Lagrangian written in components
\beqas
{\cal L}_{FOG}  & = & \tilde{\epsilon}^{\mu \nu}(X^+(\partial_\mu - \omega_\mu)e^-_\nu + X^-(\partial_\mu + \omega_\mu)e^+_\nu \\
                &   & + X \partial_\mu \omega_\nu) - (e){\cal V}(X^+X^-;X) \\
{\cal L}_{kin}  & = & \frac{i}{\sqrt{2}} f(X)\left[ -e_0^+(\chid_0\plr{1}\chi_0)
                                                     +e_0^-(\chid_1\plr{1}\chi_1)\right.\\
           & & \hspace{19mm}\left.                   +e_1^+(\chid_0\plr{0}\chi_0)
                                                     -e_1^-(\chid_1\plr{0}\chi_1)
                                               \right]\\
{\cal L}_{SI}   & = & (e_0^- e_1^+ - e_0^+e_1^-) h(X) g(\chid_1 \chi_0 + \chid_0 \chi_1)
\eeqas
shows that there do not occur any time (i.e. $x^0$) derivatives of the
$\ol{q}^i$ and thus the $\ol{p}_i$ are constrained to zero, $\ol{p}_i
\approx 0$, where $\approx$ means weakly equal to zero.

Because the kinetic term for the fermions is of first order in the
derivatives, the fermion momenta $P_\al := \pld{\cal L}{\dot{Q}^\al}$ give
rise to primary constraints
\beqa\label{Phi0}
\Phi_0 & = & P_0 + \frac{i}{\sqrt{2}}f(p_1) q^3 Q^2 \approx 0\\\label{Phi1}
\Phi_1 & = & P_1 - \frac{i}{\sqrt{2}}f(p_1) q^2 Q^3 \approx 0\\\label{Phi2}
\Phi_2 & = & P_2 + \frac{i}{\sqrt{2}}f(p_1) q^3 Q^0 \approx 0\\\label{Phi3}
\Phi_3 & = & P_3 - \frac{i}{\sqrt{2}}f(p_1) q^2 Q^1 \approx 0
\eeqa
They have a non vanishing Poisson bracket\footnote{$\spoiss{.}{.}$
  denotes the graded Poisson bracket. For the definition c.f. App. A}
with each other,
\beqa\label{Calphabeta}\nonumber
C_{\al\be}(x,y) & := & \spoiss{\Phi_\al(x)}{\Phi_\be(y)} \\
                & = & i\sqrt{2}f(X)
                      \left(
                      \begin{matrix}
                      0      & 0     & -e_1^+ & 0     \\
                      0      & 0     & 0      & e_1^- \\
                      -e_1^+ & 0     & 0      & 0     \\
                      0      & e_1^- & 0      & 0
                      \end{matrix}
                      \right)\de(x-y)
\eeqa
and thus are \textbf{of second class}, according to Dirac's original
classification of constraints \cite{Dirac:1996}. The $\Phi_\al$,
however, are independent of the $\ol{q}^i$ and thus commute with the
$\ol{p}_i$.

Having computed all the momenta, we obtain the \textbf{Hamiltonian
  density}
\beqa\nonumber
\cal H         & =  & \dot{Q}^\al P_\al + p_i \dot{q}^i - \cal L \\\label{wholehamiltonian}
               & =: & {\cal H}_{FOG} + {\cal H}_{kin} + {\cal H}_{SI} \\\nonumber
{\cal H}_{FOG} & = & X^+ (\partial_1 - \om_1)e_0^- + X^-(\partial_1 + \om_1)e_0^+ + X\partial_1 \om_0 + (e){\cal V}\\\nonumber
               &   & +(X^+e_1^- - X^-e_1^+)\om_0\\\nonumber
{\cal H}_{kin} & = & \frac{i}{\sqrt{2}}f(X)\left[e_0^+(\chid_0 \lrpd{1} \chi_0) - e_0^-(\chid_1 \lrpd{1} \chi_1) \right]\\\nonumber
{\cal H}_{SI}  & = & -(e)h(X)g(\chib\chi)
\eeqa

To deal with the second class constraints, we pass to the \textbf{Dirac bracket}
\cite{Dirac:1996,Henneaux:1992}
\beq\label{diracbracket}
\dirac{f(x)}{g(y)} := \spoiss{f}{g} - \int dzdw \; \spoiss{f(x)}{\Phi_\al(z)} C^{\al\be}(z,w) \spoiss{\Phi_\be(w)}{g(y)}
\eeq
with $C^{\al\be}(x,y)$ being the inverse of the matrix-valued
distribution
\footnote{The inverse $C^{\al\be}(x,y)$ of a matrix valued distribution $C_{\al\be}(x,y)$ is defined such that
          $\int dy \left(\int dx \varphi(x) C_{\al\ga}(x,y)\right)\left(\int dz \psi(z) C^{\ga\be}(y,z)\right) = 
           \int dx \varphi(x) \int dz \psi(z) \delta_\al^\be \delta(x-z)$ for all test functions $\varphi,\psi$.
          When $C_{\al\be}(x,y) = C_{\al\be}(x)\delta(x-y)$, then the inverse is $C^{\al\be}(x,y)=C^{\al\be}(x)\delta(x-y)$ 
          with $C_{\al\ga}(x)C^{\ga\be}(x)=\delta_\al^\be \; \forall x$}
\eqref{Calphabeta}.

Demanding that the primary first class constraints should not change
during time evolution, i.e.\footnote{Henceforth, a prime in a Dirac or Poisson bracket means
  evaluation of the function at a point $x'$.}
$G_i := \dot{\ol{p}}_i = \dirac{\ol{p}_i}{{\cal H}'} = \spoiss{\ol{p}_i}{{\cal
    H}'} \approx 0$, leads us to \textbf{secondary constraints}
\beqa                                                                                         \label{G1}
G_1 & = & G_1^g                                                                             \\\label{G2}
G_2 & = & G_2^g + \frac{i}{\sqrt{2}}f(X)(\chid_1 \lrpd{1} \chi_1) + e_1^+ h(X) g(\chib\chi) \\\label{G3}
G_3 & = & G_3^g - \frac{i}{\sqrt{2}}f(X)(\chid_0 \lrpd{1} \chi_0) - e_1^- h(X) g(\chib\chi)
\eeqa
with
\beqas
G_1^g & = & \partial_1 X + X^- e_1^+ - X^+ e_1^-                                            \\
G_2^g & = & \partial_1 X^+ + \om_1 X^+ - e_1^+ \cal V                                       \\
G_3^g & = & \partial_1 X^- - \om_1 X^- + e_1^- \cal V
\eeqas
The Hamiltonian density now turns out to be constrained to zero, as
expected for a \textbf{generally covariant system}\footnote{Up to a
  boundary term $\int\limits_{\partial \cal M} p_i \ol{q}^i$ coming
  from ${\cal L}_{FOG}$.}  \cite{Henneaux:1992}.
\beq\label{constrainedhamiltonian}
{\cal H} = -\ol{q}^i G_i
\eeq
This already lets us to expect that the $G_i$ are related to the
generators of the three gauge symmetries in the system, namely local
Lorentz symmetry and the two diffeomorphisms.

The secondary constraints commute with the $\ol{p}_i$ because both the
$G_i$ and $\Phi_\al$ are independent of the $\ol{q}^i$. They also
trivially commute with the primary second class constraints,
$\dirac{\Phi_\al}{G_j'} = 0$, because of the definition of the Dirac
bracket. For the same reason the $\Phi_\al$ do not give rise to new
secondary constraints.

\subsection{Algebra of the secondary constraints}

\textbf{Dirac conjectured} \cite{Dirac:1996} that every first class
constraint generates a gauge symmetry. The proof of this conjecture is
possible in a very general setting \cite{Gitman:1990}, but some
additional assumptions (see paragraph 3.3.2 of \cite{Henneaux:1992})
to rule out ``pathological'' examples make it easier.  These
assumptions are fulfilled in our case, because 1. every constraint
belongs to a well defined generation; 2.  the Dirac bracket ensures
that the primary second class constraints do not generate new ones
and, as will be seen below, the secondary constraints are first class
and there are no ternary constraints; and 3. every primary first class
constraint $\ol{p}_i = 0$ generates one $G_i$.

To show that the system doesn't admit any ternary constraints, it is
sufficient to show that the algebra of secondary constraints closes,
i.e. $\dirac{G_i}{G_j '} = \left.C_{ij}\right.^k (x)G_k\;\de(x-x')$, and thus
the secondary constraints are preserved under the time evolution
generated through the Dirac bracket,
\beqs
\dot{G}_i = \dirac{G_i}{\left.{\cal H}\right.' } = - \left.\ol{q}^j\right. ' \dirac{G_i}{G_j'} \approx 0
\eeqs

To calculate all the Dirac brackets, one first needs the Poisson
brackets $\spoiss{\Phi_\al}{G_j}$. They are rather lengthy and thus
listed in Appendix \ref{app:B}. The resulting \textbf{algebra} then
reads
\beqa\label{GiGi}
\dirac{G_i}{G_i '} & = & 0 \hspace{5mm} i = 1,2,3 \\\label{G1G2}
\dirac{G_1}{G_2 '} & = & - G_2 \,\de \\\label{G1G3}
\dirac{G_1}{G_3 '} & = & G_3 \,\de \\\label{G2G3}
\dirac{G_2}{G_3 '} & = & \left[ - \sum\limits_{i=1}^3 \td{{\cal V}}{p_i} G_i + \left(g h' - \frac{h}{f} f'g'\cdot (\chib\chi)\right)G_1 \right]\de
\eeqa

We'd like to comment on some technical points.  Only obtaining
\eqref{G2G3} needs some care, the others brackets are rather
straightforward, using the Poisson structure of our phase space
\eqref{poisson}. However, one should keep in mind that the $Q^\al$ are
anti commuting.  The tricky part of \eqref{G2G3} is actually not the
$C^{\al\be}$-term in the Dirac bracket, but the bracket
$\spoiss{G_2[\varphi]}{G_3[\psi]}$ and therein the integrations by
part during calculation, which have to be performed using smeared
constraints, i.e.
\beqs
G_i[\varphi] = \int dx \; \varphi(x) G_i(x)
\eeqs
The bracket itself reads with \eqref{G2}, \eqref{G3}
\beqa\nonumber
\spoiss{G_2[\varphi]}{G_3[\psi]} & = & \iint dx dz \varphi(x) \psi(z) \left( \spoiss{G_2^{old}(x)}{G_3^{old}(z)}
                                       + \spoiss{q^3(x)h(x)g(x)}{G_3^{old}(z)}\right.\\\label{sG2G3}
                                 &   & \hspace{37mm} \left.- \spoiss{G_2^{old}(x)}{q²(z)h(z)g(z)}\right)
\eeqa
Here we denote with $G^{old}$ the constraints with $h=0$,
\beqas
G_1^{old} & = & G_1^g = G_1 \\
G_2^{old} & = & G_2^g + \frac{i}{\sqrt{2}}f(X)(\chid_1 \lrpd{1} \chi_1)\\
G_3^{old} & = & G_3^g - \frac{i}{\sqrt{2}}f(X)(\chid_0 \lrpd{1} \chi_0)
\eeqas
and
\beqs
\spoiss{G_2^{old}(x)}{G_3^{old}(z)} = - \sum\limits_{i=1}^3 \td{{\cal V}}{p_i} G_i^{g}
\eeqs
The tricky parts are the second and third bracket in \eqref{sG2G3}
\footnote{The points in space where the functions are evaluated are denoted by subscript here, e.g. $h_x(p_1) := h(p_1(x))$}:
\beqas
\spoiss{(q³h(p_1)g(\chib\chi))[\varphi]}{G_3^{old}[\psi]}
& = & \iint dx dz \varphi_x \psi_z g_x(\chib\chi) \poiss{q^3_xh_x(p_1)}{G_{3,z}^g} \\
& = & \iint dx dz \varphi_x \psi_z g_x(\chib\chi) \left[ (\partial_z \de(x-z))h_x(p_1) \right.\\
&   & \hspace{15mm}\left.- (q^1h-q^3p_3 h' -q^2p_2 U h)_x \;\de(x-z)  \right]\\
\spoiss{G_2^{old}[\varphi]}{(q²h(p_1)g(\chib\chi))[\psi]}
& = & \iint dx dz \varphi_x \psi_z g_z(\chib\chi) \poiss{G_{2,x}^g}{q^2_z h_z(p_1)} \\
& = & \iint dx dz \varphi_x \psi_z g_z(\chib\chi) \left[ (-\partial_x \de(x-z))h_z(p_1) \right.\\
&   & \hspace{15mm}\left.- (q^1h-q^2p_2 h' -q^3p_3 U h)_x\;\de(x-z)  \right]
\eeqas
\beqas
& \Rightarrow & \hspace{12mm} \spoiss{(q³h(p_1)g(\chib\chi))[\varphi]}{G_3^{old}[\psi]} -
\spoiss{G_2^{old}[\varphi]}{(q²h(p_1)g(\chib\chi))[\psi]} \\
& & \hspace{3mm} = \hspace{5mm} \iint dx dz \varphi_x \psi_z [(\underbrace{(\partial_z \de(x-z))}\limits_{= -\partial_x \de(x-z)} g_x(\chib\chi) h_x(p_1) + (\partial_x \de(x-z)) g_z(\chib\chi) h_z(p_1)) \\
& & \hspace{35mm} \left. -g (q^2p_2-q³p_3)(h'(p_1)-U(p_1) h(p_1)) \de(x-z)\right]\\
& & \stackrel{int.p.p.}{=} \iint dx dz \varphi_x \psi_z \de(x-z)[ \partial_x(hg) - g(q²p_2-q³p_3)(h'-Uh) ]
\eeqas
Thus we get for the graded Poisson bracket of $G_2$ and $G_3$
\beqas\nonumber
\spoiss{G_2}{G_3 '} & = & \left[ -\td{{\cal V}}{p_i} G_i^g + \frac{i}{\sqrt{2}} f' [p_3(Q³\lrpd{x}Q^1)-p_2(Q²\lrpd{x}Q^0)]\right. \\
                    &   & \hspace{5mm}\left. + \partial_x(hg) - g(q²p_2-q³p_3)(h'-Uh)\right]\delta
\eeqas

The $C^{\al\be}$-terms of the Dirac bracket are (with $\partial_xg = g'\partial_x(\chib\chi)$, $\partial_x f = f' \partial_x p_1$ and $p_2q²-p_3q³-\partial_x p_1 = -G_1$)
\beqa\nonumber
-\frac{i}{\sqrt{2}}[f'+Uf][p_3(Q³\lrpd{x}Q^1)-p_2(Q²\lrpd{x}Q^0)] - \frac{h}{f}f'g'\cdot(\chib\chi) G_1  - h(\partial_x g)
\eeqa

With these results, we obtain (omitting the $\de(x-x')$)
\beqa\nonumber
\dirac{G_2}{G_3 '} & = & -\td{\cal V}{p_1}G_1 
                         -\td{\cal V}{p_2}\left(G_2^g + \frac{i}{\sqrt{2}} f (Q³\lrpd{x}Q^1) + q³hg\right)\\\nonumber
                   &   & -\td{\cal V}{p_3}\left(G_3^g - \frac{i}{\sqrt{2}} f (Q³\lrpd{x}Q^1) - q²hg\right)\\\nonumber
                   &   & +\underbrace{\partial_x(hg)-h(\partial_x g) -gh'(\partial_x p_1)}\limits_{=0}
                         + g h' G_1\\\nonumber
                   &   & - \frac{h}{f}f' g'\cdot (\chib\chi)G_1\\\nonumber
                   & = & - \td{\cal V}{p_i}G_i + (gh' - \frac{h}{f}f' g'\cdot (\chib\chi))G_1
\eeqa

\section{Relation to the Conformal algebra}\label{sec:witt}

As first noted in \cite{Katanaev:1994qf}, certain linear
combinations of the $G_i$ fulfil the Witt algebra. In that work FOG
coupled to scalar matter was considered. However the same result holds
in our case. New generators $G = G_1$, $H_{0/1} = q^1 G_1 \mp q^2 G_2
+ q^3 G_3$ fulfil an algebra (with $\de' = \pd{\de(x-x')}{x'}$)

\beq
\begin{array}{ll}
	\dirac{G}{G'}  = 0       & \qquad \dirac{H_i}{H_i'} = (H_1 + H_1')\de' \\
	\dirac{G}{H_i} = -G \de' & \qquad \dirac{H_0}{H_1'} = (H_0 + H_0')\de' 
\end{array}
\eeq

Some Dirac brackets needed for calculating this algebra are listed in
App. \ref{app:C}. Fourier transforming the light cone combinations
$H^\pm = H_0\pm H_1$ according to $H^+(x) = \int \frac{dk}{2\pi} L_k
e^{ikx}$; $H^-(x) = \int \frac{dk}{2\pi} \overline{L}_k e^{ikx}$ shows
that the $L_k$ (and equally $\ol{L}_k$) fulfil the classical Virasoro
algebra

\beq\label{witt}
\dirac{L_k}{L_m} = i(k-m) L_{k+m}
\eeq

\section{Discussion \& Outlook}\label{sec:conclusions}

From \eqref{GiGi} - \eqref{G2G3} its clear that the algebra of
secondary constraints closes with delta functions. This implies the
absence of ternary constraints.  The $G_i$ on-shell generate three
gauge symmetries, namely one local Lorentz symmetry ($G_1$) and
two diffeomorphisms ($G_2$, $G_3$), which can be seen from
\beqa
 \dirac{G_1}{X^\pm} & = & \mp X^\pm \de \\
 \dirac{G_{2/3}}{X} & = & \pm X^\pm \de
\eeqa
by comparing with the transformation property of $X^\pm$ under Lorentz
transformations and the Lorentz scalar $X$ under diffeomorphisms.

The nontrivial part of the algebra of first class constraints still
closes like in the case of a compact Lie groups,
$[\de_A,\de_B]={f^C}_{AB}(x)\de_C$, but rather with structure
functions than with constants.  This is especially important when
considering BRST symmetry, because the homologic perturbation series
for the BRST charge then terminates at Yang-Mills level
\cite{Grumiller:2006}.

The second term in \eqref{G2G3} deserves some remarks: First, it
vanishes for minimal coupling, i.e. for $h=f=const.$ Second, if $h
\propto f$, it becomes proportional to $f'(g-g'\cdot\chib\chi)G_1\de =
- f'\la(\chib\chi)²G_1\de$. Thus a mass term $m\chib\chi$ doesn't
change the constraint algebra at all.  Third, it doesn't contain
derivatives of $\chi$. This is different from the case of scalar
matter (see eq. E.31 in \cite{Grumiller:2001ea}), where the additional
contribution to $\poiss{G_2}{G_3}$ is proportional to
$\frac{f'}{f}{\cal L}_{scalar}$.

Boundary contributions both to the dilaton and the fermionic action
have been omitted this work. Dilaton theories with boundaries (but
without matter) have been considered recently in detail
\cite{Bergamin:2005pg}, with the result that a consistent variational
principle can be defined. We don't expect problems from the interplay
of fermion boundary terms and gravitational ones. Nevertheless this
point still has to be worked out.

As noted in Sec. \ref{sec:intro}, another motivation for this work
stems from the recent paper by Frolov, Kristjánsson and Thorlacius
\cite{Frolov:2005ps}, who considered a two-dimensional Schwinger model
on a curved background manifold to investigate the effect of
pair-production on the global structure of black hole space times. To
this end they used the quantum equivalence of the Schwinger model in
1+1 dimensions and the Sine-Gordon model found by Coleman, Jackiw \&
Susskind \cite{Coleman:1975pw} to do calculations on the Sine-Gordon
side. It is an interesting question whether Bosonisation still shows
up in a quantum theory with dynamical gravitational background. For
matter less generalised dilaton theories \eqref{dilatonaction} and for
ones coupled to scalar fields an exact path integral quantisation of
the geometric sector is known \cite{Kummer:1996hy} and gives rise to a
nonlocal vertices effective theory which turn out to be interesting
intermediary states like Virtual Black Holes \cite{Grumiller:2000ah}.
A similar analysis for our case is in preparation
\cite{Grumiller:2006} and will shed some light on the question posted
above.

\acknowledgments

This text is a contribution to the proceedings of the International
V.A. Fock School for Advances in Physics (IFSAP-2005) held in St.
Petersburg in November this year. The author would like to thank the
organisers of the school and especially Prof. Victor Novozhilov for
his sacrificial work and for the opportunity of giving a talk on the
results presented here. Special thanks deserves Dmitri Vassilevich for
recommending me for a UNESCO short term fellowship.

The results presented here are part of my diploma thesis, and I am
very grateful to my advisors, Daniel Grumiller and Prof. Gerd
Rudolph, for their support of my studies, and in particular to Daniel
\& Dmitri for many discussions on physics and two-dimensional gravity.

Last but not least I want to thank Prof. Wolfgang Kummer for the
invitation to talk in Vienna in October and Luzi Bergamin for
financial support.

\begin{appendix}

\section{Conventions}\label{app:A}

For the Levi-Civit{\'a} symbols both in tangent space $\epsilon^{ab}$and
on the manifold $\tilde{\epsilon}^{\mu\nu}$, we fix $\epsilon^{01} :=
+1$. This is necessary to retain $\epsilon^{\mu\nu}=e_a^\mu e_b^\nu
\epsilon^{ab}$, with $\epsilon^{\mu\nu}$ now being the Levi-Civit{\'a}
tensor. In the tangent space we use light cone coordinates $X^\pm =
\frac{1}{\sqrt{2}}(X^0 \pm X^1)$, and thus $\eps^{ab}_{LC} =
-\eps^{ab}$. For the square root of the determinant of the metric we
sometimes write $(e):=e_0^- e_1^+ - e_0^+e_1^- = -\det({e_\mu}^a) =
\sqrt{-\det(g_{\mu\nu})}$, whereas the volume 2-form is $\epsilon =
-(e)d^2x = \det({e_\mu}^a)dx^0 \wedge dx^1$. The Hodge star is defined
as in \cite{Grumiller:2002nm}. The two-sided derivative is
$a\overleftrightarrow{\mathrm{d}}b := a (\mathrm{d} b) -
(\mathrm{d}a)b$

Our \textbf{Dirac matrices} are
\beqs
\begin{array}{ll}
\ga^0 = \ \;\left(
\begin{array}{rr}
	0 & 1 \\
	1 & 0
\end{array}\right)
& \ \ \ 
\ga^1 = \left(
\begin{array}{rr}
	0 & 1 \\
	-1 & 0
\end{array}\right) \\
\ga^+ = \left(
\begin{array}{rr}
	0 & \sqrt{2} \\
	0 & 0
\end{array}\right)
& \ \ \ 
\ga^- = \left(
\begin{array}{rr}
	0 & 0 \\
	\sqrt{2} & 0
\end{array}\right)
\end{array}
\eeqs
The analogue of $\ga_5$ is defined as $\ga_* := \ga^0\ga^1 =
\frac{1}{2}[\ga^0,\ga^1] $.
Because our fermion fields are Grassmann variables, throughout this
article we use the \textbf{graded Poisson bracket} defined as
\cite{Henneaux:1992} ($\partial^L$ is the usual left derivative)
\beqas
\spoiss{F}{G} & = & \int dz \left[ \pd{F}{q^i(z)} \pd{G}{p_i(z)} - \pd{F}{p_i(z)} \pd{G}{q^i(z)} \right] \\ 
              &   & + (-)^{\eps(F)} \left[ \pld{F}{Q^\al(z)} \pld{G}{P_\al(z)} - \pld{F}{P_\al(z)} \pld{G}{Q^\al(z)} \right]
\eeqas
with $(q,p)$ and $(Q,P)$ being a set of bosonic ($\eps(q)=\eps(p)=0$)
and fermionic ($\eps(Q)=\eps(P) =1$), and $\eps(F)$ the Grassmann
parity of $F$. Its main
properties used here are
\beqas
\spoiss{F}{G} & = & (-)^{\eps(F)\eps(G) + 1} \spoiss{G}{F} \\
\spoiss{F}{G_1 G_2} & = & \spoiss{F}{G_1} G_2 + (-)^{\eps(F)\eps(G_1)} G_1 \spoiss{F}{G_2}
\eeqas
All these properties carry over to the corresponding Dirac bracket
defined by \eqref{diracbracket}.

\section{Brackets of the secondary with the second class constraints}\label{app:B}

To calculate the Dirac brackets, we need all the graded Poisson
brackets of the $G_i$ with the $\Phi_\al$. They are easily obtained by
using the algebraic properties of the graded Poisson bracket (see App.
\ref{app:A}).
\beqas
\spoiss{G_1}{\Phi_0'} & = & -\frac{i}{\sqrt{2}} f e_1^+ \chid_0 \;\de(x-x')\\
\spoiss{G_1}{\Phi_2'} & = & -\frac{i}{\sqrt{2}} f e_1^+ \chi_0  \;\de(x-x')\\
\spoiss{G_1}{\Phi_1'} & = & -\frac{i}{\sqrt{2}} f e_1^- \chid_1 \;\de(x-x')\\
\spoiss{G_1}{\Phi_3'} & = & -\frac{i}{\sqrt{2}} f e_1^- \chi_1  \;\de(x-x')\\
\spoiss{G_2}{\Phi_0'} & = & \frac{i}{\sqrt{2}} \left[ f'+Uf \right] X^+ e_1^+ \chid_0\;\de(x-x')\\
                      &   & -\; e_1^+ h g' \chid_1 \;\de(x-x')\\
\spoiss{G_2}{\Phi_2'} & = & \frac{i}{\sqrt{2}} \left[ f'+Uf \right] X^+ e_1^+ \chi_0\;\de(x-x')\\
                      &   & +\; e_1^+ h g' \chi_1  \;\de(x-x')
\eeqas

However one must be careful with integrating by parts the derivatives
of the delta distributions. This is most easily done by smearing the
fields with appropriate test functions
\footnote{After integrating by parts one obtains
\beqs
\int dx \varphi(x)\left[ f(x) - f(y) \right]\partial_x\de(x-y) = - \int dx \varphi(x) (\partial_x f(x)) \delta(x-y)
\eeqs},
\beqas
\spoiss{G_2}{\Phi_1'} & = & \frac{i}{\sqrt{2}} \left[ \chid_1(\om_1 f - X^+ e_1^- f' - X^- e_1^+ U f)
                            + 2 (\partial_x \chid_1)f  + (\partial_x f)\chid_1 \right] \de(x-x')\\
                      &   & -\; e_1^+ h g' \chid_0 \;\de(x-x')\\
\spoiss{G_2}{\Phi_3'} & = & \frac{i}{\sqrt{2}} \left[ \chi_1(\om_1 f - X^+ e_1^- f' - X^- e_1^+ U f)
                            + 2 (\partial_x \chi_1)f  + (\partial_x f)\chi_1 \right] \de(x-x')\\
                      &   & +\; e_1^+ h g' \chi_0 \;\de(x-x')\\
\spoiss{G_3}{\Phi_0'} & = & \frac{i}{\sqrt{2}} \left[ \chid_0(\om_1 f - X^- e_1^+ f' - X^+ e_1^- U f)
                            - 2 (\partial_x \chid_0)f  - (\partial_x f)\chid_0 \right] \de(x-x')\\
                      &   & +\; e_1^- h g' \chid_1 \;\de(x-x')\\\nonumber
\spoiss{G_3}{\Phi_2'} & = & \frac{i}{\sqrt{2}} \left[ \chi_0(\om_1 f - X^- e_1^+ f' - X^+ e_1^- U f)
                            - 2 (\partial_x \chi_0)f  - (\partial_x f)\chi_0 \right] \de(x-x')\\
                      &   & -\; e_1^- h g' \chi_1 \;\de(x-x')\\\nonumber
\spoiss{G_3}{\Phi_1'} & = & \frac{i}{\sqrt{2}} \left[ f'+Uf \right] X^- e_1^- \chid_1\;\de(x-x')\\
                      &   & +\; e_1^- h g' \chid_0 \;\de(x-x')\\\nonumber
\spoiss{G_3}{\Phi_3'} & = & \frac{i}{\sqrt{2}} \left[ f'+Uf \right] X^- e_1^- \chi_1\;\de(x-x')\\
                      &   & -\; e_1^- h g' \chi_0 \;\de(x-x')
\eeqas

\section{Dirac Brackets needed for eq. \eqref{witt}}\label{app:C}

\beqas
	\dirac{G_1}{q^1} & = & -\partial_1 \de \\
	\dirac{G_1}{q^2} & = & q^2 \de \\
	\dirac{G_1}{q^3} & = & -q^3 \de \\
	\dirac{G_2}{q^1} & = & q^3\left[ \pd{\cal V}{p_1} - \left(h' g - \frac{f'}{f}h g' \cdot (\chib\chi)\right) \right] \de\\ 
	\dirac{G_2}{q^2} & = & - \left[ \partial_1 + q^1 - q^3\pd{\cal V}{p_2}\right]\de \\
	\dirac{G_2}{q^3} & = & q^3 \pd{\cal V}{p_3} \de
\eeqas
\beqas
	\dirac{G_3}{q^1} & = & -q^2\left[ \pd{\cal V}{p_1} - \left(h' g - \frac{f'}{f}h g' \cdot (\chib\chi)\right) \right] \de \\
	\dirac{G_3}{q^2} & = & -q^2 \pd{\cal V}{p_2} \de \\
	\dirac{G_3}{q^3} & = & - \left[ \partial_1 - q^1 + q^2\pd{\cal V}{p_3}\right]\de
\eeqas
\beqas
  \dirac{q^iG_i}{q^iG_i} & = & -(\partial_1\de) q^i G_i \ \ \mathrm{(no\ summation)} \\
  \dirac{q^1G_1}{q^2G_2} & = & -q^2q^3\left[ \pd{\cal V}{p_1} - \left(h' g - \frac{f'}{f}h g' \cdot (\chib\chi)\right) \right] G_1 \de
                           = - \dirac{q^1G_1}{q^3G_3} = \dirac{q^2G_2}{q^3G_3}
\eeqas

\end{appendix}

\providecommand{\href}[2]{#2}\begingroup\raggedright\endgroup


\end{document}